\documentclass[aps,pra, reprint,superscriptaddress]{revtex4-2}
\usepackage{amsmath}
\usepackage{amssymb}
\usepackage{xcolor}
\usepackage{graphicx}
\usepackage{hyperref}
\usepackage{braket}
\usepackage{color}
\usepackage{ulem}
\usepackage{mathptmx}
\usepackage{dcolumn}
\usepackage{bm}
\usepackage{here}

\makeatletter
\newcommand{\subscripts}[3]{%
  \@mathmeasure\z@\displaystyle{#2}%
  \global\setbox\@ne\vbox to\ht\z@{}\dp\@ne\dp\z@
  \setbox\tw@\box\@ne
  \@mathmeasure4\displaystyle{\copy\tw@_{#1}}%
  \@mathmeasure6\displaystyle{{#2}_{#3}}%
  \dimen@-\wd6 \advance\dimen@\wd4 \advance\dimen@\wd\z@
  \hbox to\dimen@{}\mathop{\kern-\dimen@\box4\box6}%
}

\makeatother
\begin{document}

\title{Non-Clifford gate on optical qubits by nonlinear feedforward}


\author{Shunya Konno}
\affiliation{Department of Applied Physics, School of Engineering, The University of Tokyo, 7-3-1 Hongo, Bunkyo-ku, Tokyo 113-8656, Japan}
\author{Warit Asavanant}
\affiliation{Department of Applied Physics, School of Engineering, The University of Tokyo, 7-3-1 Hongo, Bunkyo-ku, Tokyo 113-8656, Japan}
\author{Kosuke Fukui}
\affiliation{Department of Applied Physics, School of Engineering, The University of Tokyo, 7-3-1 Hongo, Bunkyo-ku, Tokyo 113-8656, Japan}
\author{Atsushi Sakaguchi}
\affiliation{Department of Applied Physics, School of Engineering, The University of Tokyo, 7-3-1 Hongo, Bunkyo-ku, Tokyo 113-8656, Japan}
\affiliation{Optical Quantum Computing Research Team, RIKEN Center for Quantum Computing, 2-1 Hirosawa, Wako, Saitama, 351-0198, Japan}
\author{Fumiya Hanamura}
\affiliation{Department of Applied Physics, School of Engineering, The University of Tokyo, 7-3-1 Hongo, Bunkyo-ku, Tokyo 113-8656, Japan}
\author{Petr Marek}
\affiliation{Department of Optics, Palack\'y University, 17. listopadu 1192/12, 77146 Olomouc, Czech Republic}
\author{Radim Filip}
\affiliation{Department of Optics, Palack\'y University, 17. listopadu 1192/12, 77146 Olomouc, Czech Republic}
\author{Jun-ichi Yoshikawa}
\affiliation{Department of Applied Physics, School of Engineering, The University of Tokyo, 7-3-1 Hongo, Bunkyo-ku, Tokyo 113-8656, Japan}
\author{Akira Furusawa}
\email{akiraf@ap.t.u-tokyo.ac.jp}
\affiliation{Department of Applied Physics, School of Engineering, The University of Tokyo, 7-3-1 Hongo, Bunkyo-ku, Tokyo 113-8656, Japan}
\affiliation{Optical Quantum Computing Research Team, RIKEN Center for Quantum Computing, 2-1 Hirosawa, Wako, Saitama, 351-0198, Japan}


\date{\today}

\begin{abstract}
In a continuous-variable optical system, the Gottesman-Kitaev-Preskill (GKP) qubit is a promising candidate for fault-tolerant quantum computation. To implement non-Clifford operations on GKP qubits, non-Gaussian operations are required. In this context, the implementation of a cubic phase gate by combining nonlinear feedforward with ancillary states has been widely researched.  Recently, however, it is pointed out that the cubic phase gate is not the most suitable for non-Clifford operations on GKP qubits. In this work, we show that we can achieve linear optical implementation of non-Clifford operations on GKP qubit with high fidelity by applying the nonlinear feedforward originally developed for the cubic phase gate and using a GKP-encoded ancillary state. Our work shows the versatility of nonlinear feedforward technique important for optical implementation of the fault-tolerant continuous-variable quantum computation.

\end{abstract}


\maketitle

\section{Introduction}
Quantum computation holds a key to a computational power that supersedes the current classical computers \cite{nielsen00}. Among many physical candidates, continuous-variable (CV) quantum computation using propagating optical fields has many distinctive features such as scalability. Recently, optical CV cluster states, fundamental computational resource states for one-way quantum computation \cite{PhysRevLett.86.5188,PhysRevLett.97.110501}, have been deterministically realized in a scalable fashion \cite{Asavanant373,Larsen369,PhysRevLett.112.120505} and some basic operations based on cluster states have already been demonstrated \cite{2020arXiv200611537A, Larsen2021}. 

Cluster states by themselves, however, are insufficient for universal CV quantum computation. CV computational resources can be divided into Gaussian and non-Gaussian resources, and both types are required to realize the universal quantum computer that has practical usages \cite{PhysRevLett.88.097904}. In order to realize non-Gaussian operations on the propagating optical fields, a key technology is a nonlinear feedforward, in which operations according to the results of nonlinear calculations to measured values are performed. For example, by combining nonlinear feedforward with appropriate ancillary states, we can implement one of the non-Gaussian operations called a cubic phase gate \cite{PhysRevA.93.022301}. Many experimental developments are on the way to realize the cubic phase gate \cite{PhysRevA.90.060302,8426782,Yukawa:13,PhysRevApplied.15.024024}. In particular, nonlinear feedforward has been developed using a low-latency digital field programmable gate array (FPGA) \cite{8426782}. In principle, if the cubic phase gate is realized, universal CV quantum computation can be achieved by combining the cubic phase gate with current Gaussian resources.

Moreover, fault-tolerant quantum computation is achievable by encoding a logical qubit in the CV system. Currently, the most promising candidate is the encoding into Gottesman-Kitaev-Preskill (GKP) qubits \cite{gottesman2001encoding}. By combining GKP qubits with CV cluster states, the fault-tolerant universal quantum computation can be achieved, even with finite squeezing \cite{menicucci2014fault,PhysRevX.8.021054,PhysRevA.100.010301,PhysRevLett.123.200502,yamasaki2020cost}. Experimentally, GKP states have been created in the ion-trapped system \cite{fluhmann2019encoding} and the microwave system \cite{campagne2020}, and error correction using GKP states has also been demonstrated \cite{campagne2020, 2020arXiv201009681D}. We can expect the optical realization of GKP states in a near future \cite{Vasconcelos:10, Bourassa2021blueprintscalable, PhysRevA.101.032315}. 

Regarding the operations on the GKP qubits, Clifford operations can be realized using only Gaussian operations, while the implementation of non-Clifford operations requires non-Gaussian elements. In the original GKP's paper \cite{gottesman2001encoding}, two types of methods are suggested to implement the T gate \cite{nielsen00}, a non-Clifford operation sufficient for full processing of GKP qubits. One method is to use the cubic phase gate combined with Gaussian operations. Recently, however, it was pointed out that the cubic phase gate is not the most suitable for the T gate on GKP qubits \cite{hastrup2021cubic}. This can be somewhat expected as the cubic phase gate is a gate intended for universal processing of CV systems \cite{PhysRevLett.82.1784} and not tailored for non-Clifford operations on GKP qubits. Since the wave function of the GKP state has a periodic structure, it is inferred that the ideal non-Clifford gate on the GKP state also has periodical action on the wave function. Such a gate is considered to be a highly non-Gaussian operation, and it is hard to construct it by using a single cubic phase gate, which is the lowest order non-Gaussian operation. The other method is based on a magic-state injection method. In this method, a non-Gaussian ancillary state is used together with Clifford operations and measurements. The non-Gaussian ancilla is a magic state, which can be distilled from noisy ancillary states \cite{PhysRevA.71.022316}. 

In the magic state injection method, a quantum nondemolition (QND) gate is used as a two-mode interaction gate. Traditionally, the QND gate has been demonstrated by applying nonlinear optical effects directly to the input modes in a nonlinear crystal \cite{PhysRevLett.72.214}. In this method, there is a large experimental loss because of low coupling efficiency between the input modes and an optical parametric oscillator, which enhances the nonlinear effect. In order to avoid this problem,  a measurement-induced QND gate has been widely studied and demonstrated in experiments recently \cite{PhysRevA.71.042308, yoshikawa2008demonstration, PhysRevA.98.052311}. By using offline ancillary states and feedforwards, we can apply nonlinear effects on the input modes indirectly without coupling loss. However, this implementation of the QND gate is also subject to another problem: the intrinsic noise in the measurement and feedforward due to the ancillary states, which is equal to ``quantum duty'' in the context of quantum teleportation \cite{PhysRevLett.80.869}. This noise can be suppressed by using squeezed ancillary states, but we can not be free from it because only finite squeezing is available in real experiments.  In order to prevent noise accumulation, the number of ancillary states should be reduced as much as possible. If we try to realize the T gate based on the QND gate, the whole setup requires many ancillary states, similar to the setup of the cubic phase gate based on the QND gate \cite{PhysRevA.84.053802}. This problem causes a degradation in the gate performance and error correction capability.

In this work, we propose a feasible T gate on propagating optical fields based on the magic state injection method. Our setup is composed of linear optics with nonlinear feedforwards and minimal offline ancillary states, which results in a reduction of noise from ancillary states. We analyze the performance of our scheme and find that it can work as an almost ideal T gate with high fidelity $\simeq 1$. In contrast, when using the cubic phase gate as the T gate based on GKP's original proposal,  the fidelity saturates at $\simeq 0.78$ \cite{hastrup2021cubic}. We also find that this fidelity can be improved to $\simeq 0.95$ by optimizing the gains of the cubic phase gate and other Gaussian gates. Our work shows the versatility of nonlinear feedforward important for optical realizations of quantum computing and is a crucial step towards the realization of the fault-tolerant optical universal quantum computer.

\section{Notation}
We consider an optical system with quadrature operators $\hat{x}$ and $\hat{p}$. We define $\hbar = 1$, thus $[\hat{x}, \hat{p}] = i $.  GKP quantum error-correcting code \cite{gottesman2001encoding}, which encodes a logical qubit in CV degrees of freedom, is a promising way to realize a fault-tolerant optical quantum computer. The ideal square lattice GKP qubit is defined as
\begin{eqnarray}
\ket{0_\mathrm{L}} \propto \sum_{s\in \mathbb{Z}} \ket{2s\sqrt{\pi}},~ \ket{1_\mathrm{L}} \propto \sum_{s\in \mathbb{Z}} \ket{(2s+1)\sqrt{\pi}} , \label{idealGKP} 
\end{eqnarray}
where "L" indicates logical qubit, $\ket{x}$ is the eigenstate of the quadrature operator $\hat{x}$ as $\hat{x} \ket{x} = x \ket{x}$, and $s$ takes all integers in $\mathbb{Z}$.

Ideal GKP qubits defined as Eqs.~(\ref{idealGKP}) are unphysical as they have infinite energy.  Approximate GKP states which are physically realizable can be obtained by replacing each $\hat{x}$ eigenstate in the ideal GKP states with finite squeezed vacuum weighted by a Gaussian envelope \cite{gottesman2001encoding}:

\begin{eqnarray}
\ket{0_\Delta} &\propto& \sum_{s\in \mathbb{Z}} \mathrm{e}^{-\pi (2s)^2 (2\Delta^2)/2 } \int dx~ \mathrm{e}^{-\frac{(x-2s\sqrt{\pi})^2}{2(2\Delta^2)} } \ket{x} \label{approximate0L} \\
\ket{1_\Delta} &\propto& \sum_{s\in \mathbb{Z}} \mathrm{e}^{-\pi (2s+1)^2 (2\Delta^2)/2} \int dx~ \mathrm{e}^{-\frac{(x-(2s+1)\sqrt{\pi})^2}{2(2\Delta^2)} } \ket{x},  \label{approximate1L} 
\end{eqnarray}
where $\Delta^2$ is the variance of quadrature $x$ for each peak. Squeezing level in decibel is defined as $-10\log_{10} 2\Delta^2$, which expresses the degree of approximation.

\section{Implementation of T gate}
In order to achieve universal quantum computation on GKP qubits, we have to realize the non-Clifford gate such as the T gate \cite{nielsen00}:
\begin{equation}
\hat{T} = \ket{0_\mathrm{L}} \bra{0_\mathrm{L}} + \mathrm{e}^{i \frac {\pi}{4}} \ket{1_\mathrm{L}} \bra{1_\mathrm{L}}.
\end{equation}

\begin{figure}[t]
\centering
\includegraphics[scale = 0.5]{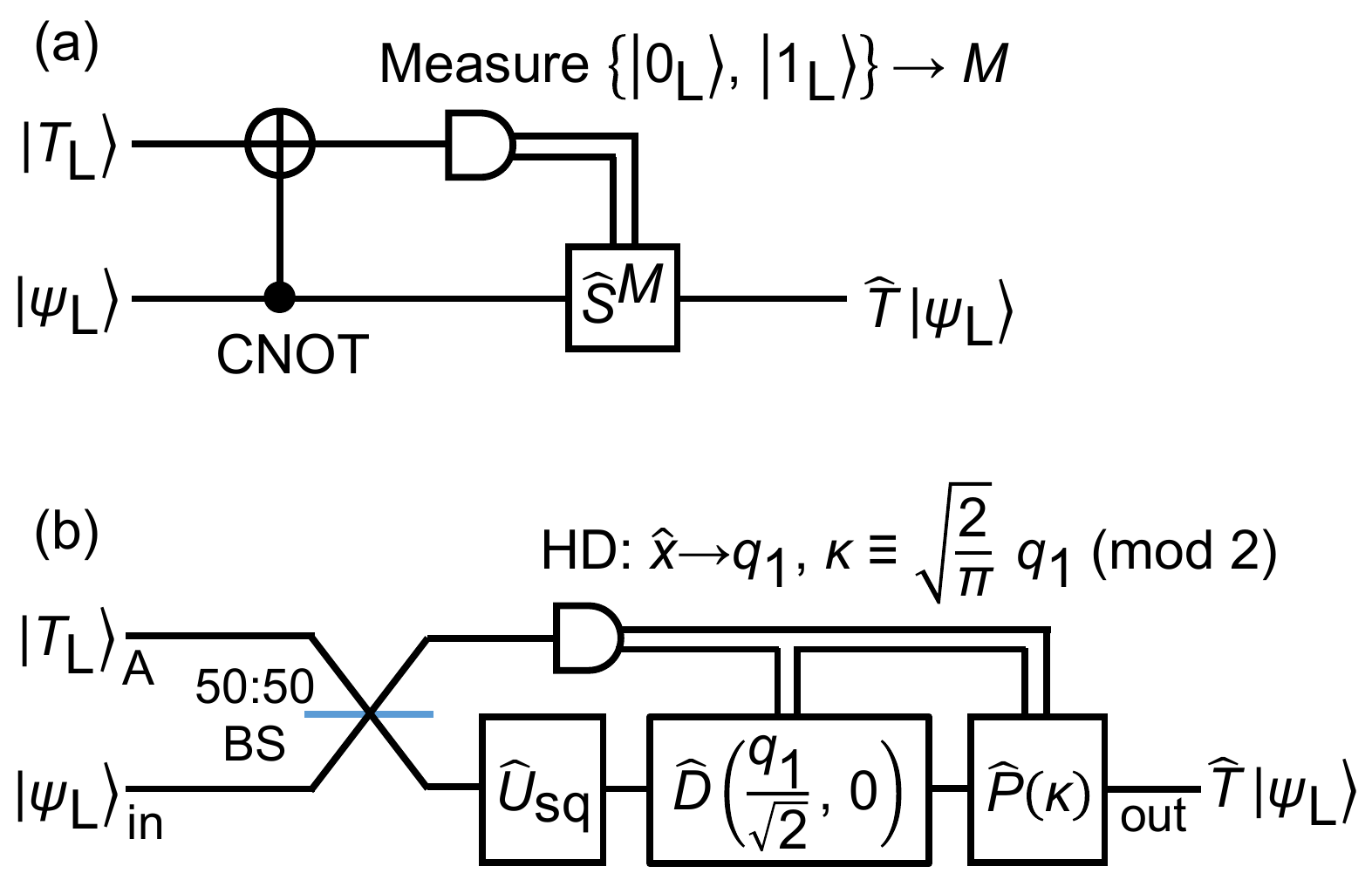}
\caption{Quantum circuits to implement the T gate. (a)Magic state injection suggested in the original GKP's paper \cite{gottesman2001encoding}. (b)The linear optical system to implement the T gate. We replace CNOT gate with a 50:50 beam splitter (BS) instead of QND gate. The measurement and the phase gate $\hat{S}$ in Fig.~1(a) are replaced by homodyne detection (HD) and shear operation $\hat{P}(\kappa)$ respectively. We add extra squeezing operation $\hat{U}_\mathrm{sq}$ and displacement operators to feedforward to fully achieve the T gate at Fig.~1(a). }
\label{fig1}
\end{figure}

We first consider ideal GKP qubits with infinite squeezing written as Eqs.~(\ref{idealGKP}). To realize the T gate for GKP qubit, a protocol using magic state injection as shown in Fig.~\ref{fig1}(a) is suggested in the original GKP paper \cite{gottesman2001encoding}. This protocol uses offline ancillary state $\ket{T_\mathrm{L}} = \frac{1}{\sqrt{2}} \left( \ket{0_\mathrm{L}} + \mathrm{e}^{i\frac{\pi}{4}} \ket{1_\mathrm{L}}\right)$, \{$\ket{0_\mathrm{L}}, \ket{1_\mathrm{L}}$\} basis measurement, controlled-NOT (CNOT) gate, and phase gate $\hat{S}= \ket{0_\mathrm{L}} \bra{0_\mathrm{L}} + i \ket{1_\mathrm{L}} \bra{1_\mathrm{L}}$. Each component has the following correspondence in the CV system. The \{$\ket{0_\mathrm{L}}, \ket{1_\mathrm{L}}$\} basis measurement is implemented by homodyne measurement, and CNOT and $\hat{S}$ operations are implemented by the QND gate $\hat{U}_\mathrm{QND} = \exp \left( i \hat{x}_1 \hat{p}_2 \right)$ and shear gate $\hat{P}(\kappa=1)$ where $\hat{P}(\kappa) =\exp \left( \frac{i}{2}\kappa \hat{x}^2 \right)$, respectively. However, the T gate constructed in this way requires a lot of resources. QND gate requires two squeezed states as ancillary states \cite{PhysRevA.71.042308, yoshikawa2008demonstration, PhysRevA.98.052311}, and shear gate requires a squeezed state \cite{PhysRevA.71.042308, PhysRevA.90.060302}.  When $\ket{T_\mathrm{L}}$ is included, the T gate demands four ancillary states in total, which makes the experimental setup complex.

In order to realize the T gate with optical beam splitter coupling instead of complicated QND interaction that demands a lot of resources \cite{PhysRevA.71.042308, yoshikawa2008demonstration, PhysRevA.98.052311}, we first propose a circuit as shown in Fig.~\ref{fig1}(b). In this circuit, the \{$\ket{0_\mathrm{L}}, \ket{1_\mathrm{L}}$\} basis measurement and the phase gate $\hat{S}$ are directly implemented by their CV correspondences mentioned above (homodyne measurement and the shear operation $\hat{P}(\kappa)$, respectively).  To achieve the T gate, we use extra squeezing $\hat{U}_{sq}$ and {displacement} operators $\hat{D} \left( \frac{q_1}{\sqrt{2}}, 0\right) $ added to feedforward. $\hat{U}_{\mathrm{sq}} = \exp \left[\frac{i}{2} (\ln \sqrt{2}) (\hat{x}\hat{p} + \hat{p}\hat{x}) \right]$ is a squeezing operator which works as $\hat{U}_{\mathrm{sq}}^\dagger \hat{x} \hat{U}_{\mathrm{sq}} = \frac{1}{\sqrt{2}} \hat{x}$ and $\hat{U}_{\mathrm{sq}}^\dagger \hat{p} \hat{U}_{\mathrm{sq}} = \sqrt{2} \hat{p}$, and $\hat{D}(x_0, p_0) = \exp \left[-i(x_0\hat{p} - p_0\hat{x}) \right]$ is a displacement operator which works as $\hat{D}^\dagger(x_0, p_0) \hat{x} \hat{D}(x_0, p_0) = \hat{x} + x_0$ and $\hat{D}^\dagger(x_0, p_0) \hat{p} \hat{D}(x_0, p_0) = \hat{p} + p_0$. We can confirm this circuit works as the T gate as follows. First, the ancillary state $\ket{T_\mathrm{L}}_{\mathrm{A}}$ and arbitrary GKP qubit input state $\ket{\psi_\mathrm{L}}_{\mathrm{in}} = a \ket{0_\mathrm{L}}_\mathrm{in} + b \ket{1_\mathrm{L}}_\mathrm{in}$ interact at the 50:50 beam splitter. The state after the 50:50 beam splitter is   
\begin{widetext}
\begin{multline}
\frac{1}{\sqrt{2}}\sum_{s, s'\in \mathbb{Z}} \left(  a \Ket{\sqrt{\frac{\pi}{2}}(-2s + 2s')}_\mathrm{A} \Ket{\sqrt{\frac{\pi}{2}}(2s + 2s')}_\mathrm{in} + b \Ket{\sqrt{\frac{\pi}{2}}(-2s + 2s'+1)}_\mathrm{A} \Ket{\sqrt{\frac{\pi}{2}}(2s + 2s'+1)}_\mathrm{in} \right. \\
\left. +\mathrm{e}^{i\frac{\pi}{4}} a \Ket{\sqrt{\frac{\pi}{2}}(-2s + 2s'-1)}_\mathrm{A} \Ket{\sqrt{\frac{\pi}{2}}(2s + 2s' +1)}_\mathrm{in}
+ \mathrm{e}^{i\frac{\pi}{4}} b \Ket{\sqrt{\frac{\pi}{2}}(-2s + 2s')}_\mathrm{A} \Ket{\sqrt{\frac{\pi}{2}}(2s + 2s'+2)}_\mathrm{in} \right). \label{eq:afterBS}
\end{multline}
\end{widetext}
Afterward, the quadrature $\hat{x}$ of mode ``A'' is measured and we obtain the measurement value $q_1$. We define $\kappa \equiv \sqrt{\frac{2}{\pi}} q_1  (\mathrm{mod}~2)$, which takes an integer value (0 or 1) as you can see from Eq.~(\ref{eq:afterBS}). The operations of feedforward are classified according to $\kappa$.  When $\kappa = 0$, the first and fourth term of Eq.~(\ref{eq:afterBS}) remain, so the state after the measurement is
\begin{eqnarray}
\sum_{s'\in \mathbb{Z}}  \left( a \ket{2s' \sqrt{2\pi} - q_1 }_\mathrm{in}
+ \mathrm{e}^{i\frac{\pi}{4}} b \ket{(2s'+1) \sqrt{2\pi} - q_1 }_\mathrm{in} \right) .
\end{eqnarray}
By applying squeezing $\hat{U}_{\mathrm{sq}}$ and displacement $\hat{D}(\frac{q_1}{\sqrt{2}}, 0)$, the output state becomes
\begin{eqnarray}
\sum_{s' \in \mathbb{Z}} \left( a \ket{2s' \sqrt{\pi} }_\mathrm{in}+ \mathrm{e}^{i\frac{\pi}{4}} b \ket{(2s'+ 1) \sqrt{\pi} }_\mathrm{in} \right) = \hat{T}\ket{\psi_\mathrm{L}}_\mathrm{in}
\end{eqnarray}
and we obtain the output of the T gate. Note that $P(\kappa)$ is an identity operator in this case because $\kappa$ is 0. 

When $\kappa = 1$, the second and third term of Eq.~(\ref{eq:afterBS}) remain, so the state after the measurement is
\begin{eqnarray}
\sum_{s' \in \mathbb{Z}} \left( b \ket{(2s'+1) \sqrt{2\pi} - q_1 }_\mathrm{in}
+ \mathrm{e}^{i\frac{\pi}{4}} a \ket{2s' \sqrt{2\pi} - q_1 }_\mathrm{in} \right) .
\end{eqnarray}
By applying squeezing $\hat{
U}_{\mathrm{sq}}$ and displacement $\hat{D}(\frac{q_1}{\sqrt{2}}, 0)$, we obtain
\begin{eqnarray}
\sum_{s' \in \mathbb{Z}} \left( b \ket{(2s'+1) \sqrt{\pi}}_\mathrm{in} + \mathrm{e}^{i \frac{\pi}{4}} a \ket{2s' \sqrt{\pi}}_\mathrm{in} \right) . 
\end{eqnarray}
Finally, we apply the shear operation $\hat{P}(\kappa = 1)$ and the output state is 
\begin{align}
\sum_{s' \in \mathbb{Z}} \left( \mathrm{e}^{i \frac{\pi}{2}} b \ket{(2s'+1) \sqrt{\pi}}_\mathrm{in} + \mathrm{e}^{i \frac{\pi}{4}} a \ket{2s' \sqrt{\pi}}_\mathrm{in} \right) = \mathrm{e}^{i \frac{\pi}{4}}  \hat{T} \ket{\psi_\mathrm{L}}_\mathrm{in} . 
\end{align}
This is the output of the T gate with an irrelevant global phase factor.

\begin{figure}[h]
\centering
\includegraphics[scale = 0.42]{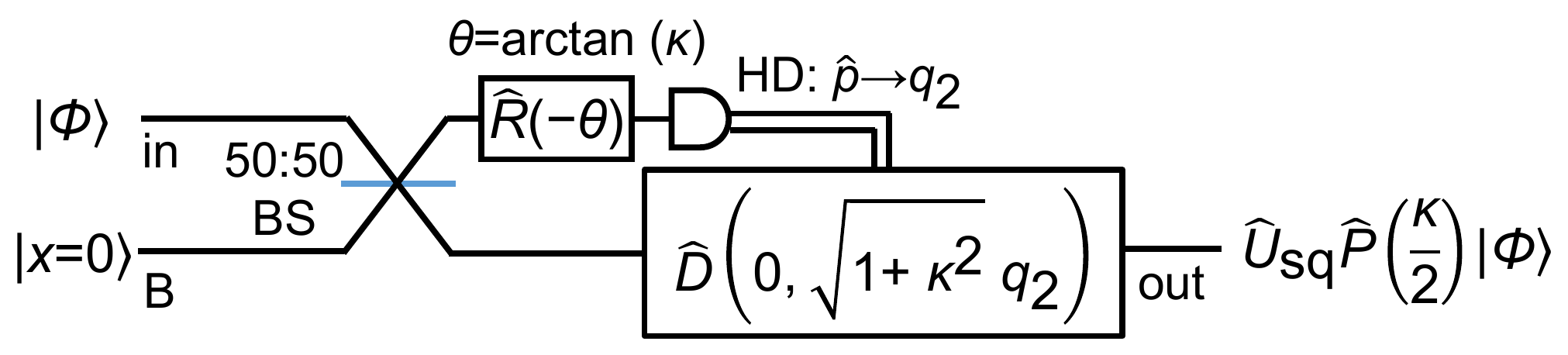}
\caption{The setup of the dynamic squeezing gate. We need only Gaussian ancillary state $\ket{x=0}_\mathrm{B}$ and linear feedforward of measured value from homodyne detection HD controling the displacement \cite{PhysRevA.90.060302}.} 
\label{fig2}
\end{figure}

\section{Actual setup}
We already proved that the T gate can be built with linear optical beam splitter coupling if the nonlinear feedforward is properly modified. The circuit shown in Fig.~\ref{fig1}(b) can be further simplified. By using the nature of unitary operations, we can change the order of operations as follows:
\begin{eqnarray}
&&\hat{P}(\kappa) \hat{D}( \frac{q_1}{\sqrt{2}},0) \hat{U}_{\mathrm{sq}} \nonumber \\
&=& \left\{ \hat{P}(\kappa) \hat{D}( \frac{q_1}{\sqrt{2}},0) \hat{P}^\dagger (\kappa) \right\}   \hat{U}_{\mathrm{sq}}  \left\{ \hat{U}_{\mathrm{sq}}^\dagger \hat{P}(\kappa) \hat{U}_{\mathrm{sq}} \right\} \nonumber \\
&=& \hat{D}( \frac{q_1}{\sqrt{2}}, \kappa \frac{q_1}{\sqrt{2}}) \hat{U}_{\mathrm{sq}} \hat{P}(\frac{\kappa}{2}) .
\end{eqnarray}
In this formula, the two consecutive operators $\hat{U}_{\mathrm{sq}} \hat{P}(\frac{\kappa}{2})$, which combine a shear gate with variable gain depending on the feedforward and a constant squeezing gate, can be realized as a dynamic squeezing gate \cite{PhysRevA.90.060302}. The setup of the dynamic squeezing gate to perform $\hat{U}_{\mathrm{sq}} \hat{P}(\frac{\kappa}{2})$ is shown in Fig.~\ref{fig2}. The input state $\ket{\Phi}_\mathrm{in}$ and the ancillary state $\ket{x=0}_\mathrm{B}$ are combined at a 50:50 beam splitter.  We implement a phase rotation $\hat{R}(-\theta) = \exp \left[ i\frac{\theta}{2} (\hat{x}^2 + \hat{p}^2) \right]$ on one mode, where $\theta$ is a variable depending on the strength of the shear gate as $\theta = \arctan (\kappa)$. Then, we measure the quadrature $\hat{p}$ and obtain the measurement value $q_2$. Finally, by performing a displacement operation $\hat{D} (0, \sqrt{1 + \kappa^2} q_2 ) $ in the remaining mode, we can obtain the output of the dynamic squeezing gate $\hat{U}_{\mathrm{sq}} \hat{P}(\frac{\kappa}{2}) \ket{\Phi}$.

\begin{figure}[t]
\centering
\includegraphics[scale = 0.37]{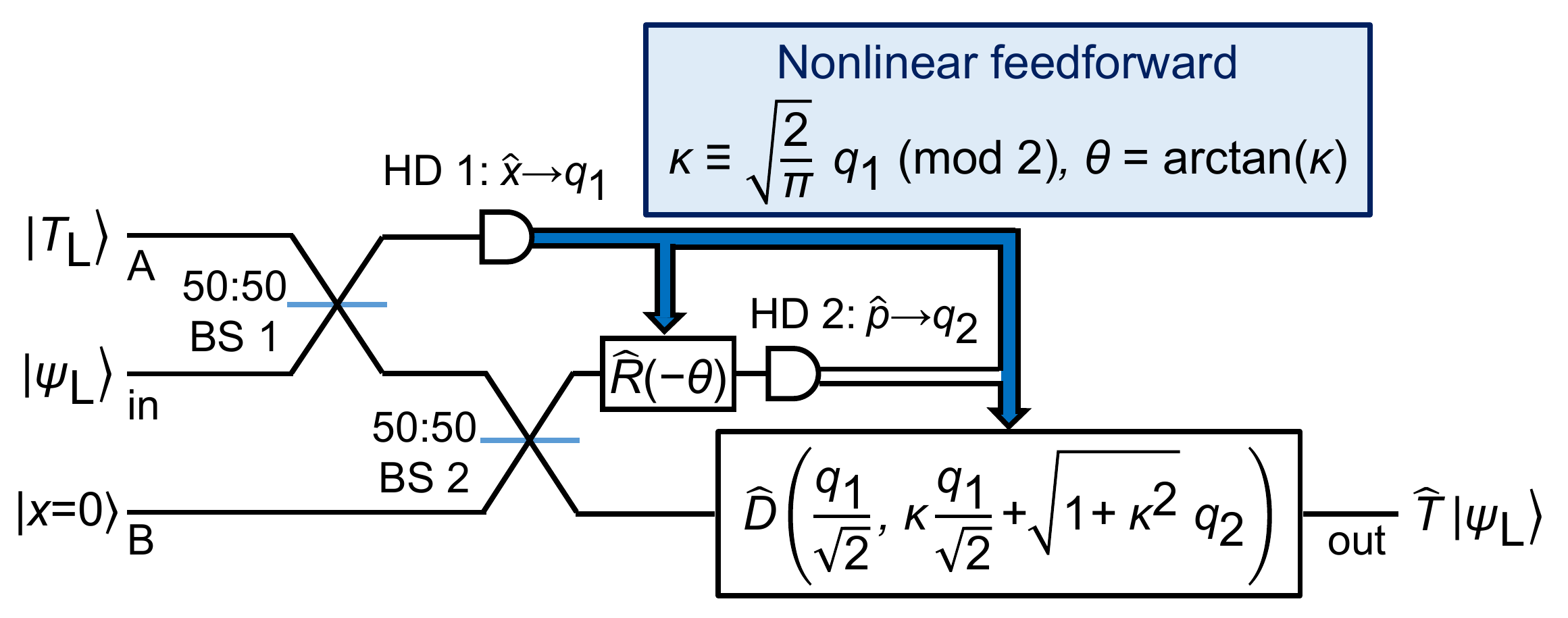}
\caption{The whole setup of the T gate. We need only two ancillary states, $\ket{T_\mathrm{L}}_\mathrm{A}$ and $\ket{x = 0}_\mathrm{B}$. The feedforward is nonlinear about the measured value from the first homodyne detection HD1. For Fig.~\ref{figfide}, we substitute ideal $\ket{T_\mathrm{L}}_\mathrm{A}$, $\ket{\psi_\mathrm{L}}_\mathrm{in}$, and $\ket{x=0}_\mathrm{B}$ by realistic $\ket{T_\Delta}_\mathrm{A}$ (Eq.~(\ref{Eq:TDelta})), $\ket{\psi_\Delta}_\mathrm{in}$ (Eq.~(\ref{Eq:inDelta})), and $\ket{\mathrm{Sq}_\sigma}_\mathrm{B}$ (Eq.~(\ref{Eq:sqsigma})). }
\label{fig3}
\end{figure}

By inserting the setup of the dynamic squeezing gate shown in Fig.~\ref{fig2}, we obtain the overall setup of the T gate shown in Fig.~\ref{fig3}. Displacement operations on the last mode are combined into one, and the value $\kappa$ of the dynamic squeezing gate is determined by the measured value $q_1$ from the first homodyne detection HD1. Both $\kappa$ and $\theta$ have a different nonlinear dependence on measurement value $q_1$. As such, nonlinear feedforward adapted to the new task, different from that of cubic phase gate \cite{PhysRevA.93.022301}, is a crucial component here. The concrete form of the feedforward operations is as follows. When $\kappa = 0$, $\theta$ is 0 thus the part of the dynamic squeezing gate becomes just a universal squeezer \cite{PhysRevA.71.042308}, and the last displacement operation is $\hat{D}(\frac{q_1}{\sqrt{2}}, q_2 ) $ where $q_2$ is the measurement value of the second homodyne detection HD2. On the other hand, when $\kappa = 1$, $\theta$ is $\frac{\pi}{4}$ and the last displacement operation is $\hat{D}(\frac{q_1}{\sqrt{2}}, \frac{q_1}{\sqrt{2}} + \sqrt{2} q_2) $.
The setup of the T gate shown in Fig.~\ref{fig3} requires only two ancillary states, non-Gaussian $\ket{T_\mathrm{L}}_\mathrm{A}$ and Gaussian $\ket{x = 0}_\mathrm{B}$, so it is much easier to construct for traveling light beams than the original setup of Fig.~\ref{fig1}(a). Most importantly, nonlinear feedforward, which is the key technology of this method has been already experimentally developed as a part of the cubic phase gate \cite{8426782}. 
Therefore, the optical T-gate can be readily constructed by applying the technology developed originally for the cubic phase gate and using $\ket{T_\mathrm{L}}$ as a non-Gaussian ancilla instead of a cubic phase state. It shows the adaptability of optical implementation based on flexible nonlinear feedforward to achieve various fault-tolerant gates for the different ancillary states.

\section{Numerical evaluation for finite squeezing}
\begin{figure}[t]
 \centering \includegraphics[angle=0, scale=0.53]{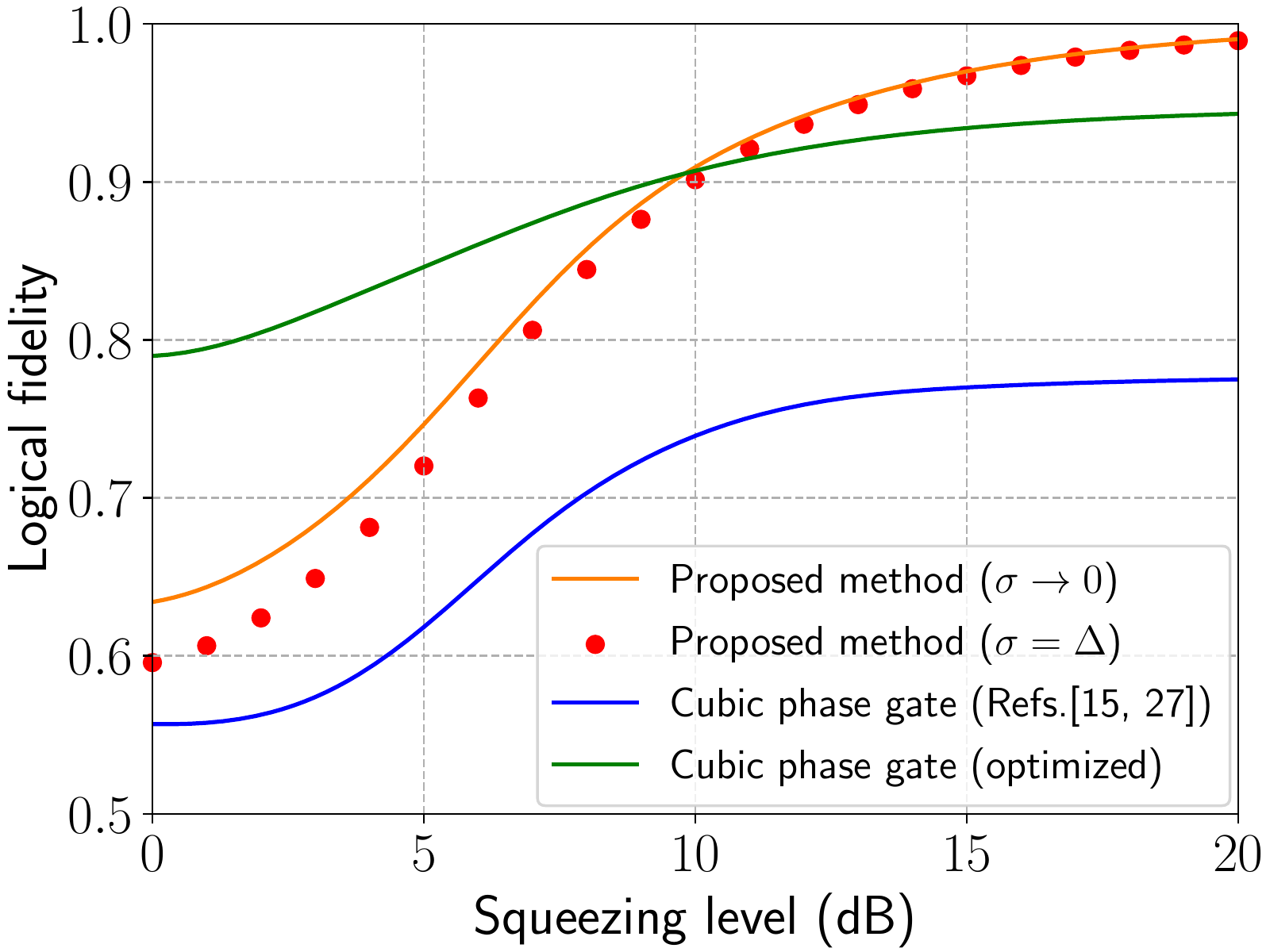} 
\caption{
The numerical evalucation of the T gate. The approximate GKP state $\ket{\psi_\Delta}_\mathrm{in}  = \frac{1}{\sqrt{2}}  \left( \ket{0_\Delta}_\mathrm{in} +  \ket{1_\Delta}_\mathrm{in} \right)$ is used as the input state, and its squeezing level is taken as the horizontal axis. The logical density matrix is obtained from the output state of the T gate, and we calculate the logical fidelity to the target state $\ket{T}_\mathcal{L} = \frac{1}{\sqrt{2}}  \left( \ket{0}_\mathcal{L} + \mathrm{e}^{i\frac{\pi}{4}} \ket{1}_\mathcal{L} \right)$. The fidelity when using our T gate setup shown in Fig.~3 is drawn by the orange line (the case the gaussian ancilla in mode ``B'' is infinitely squeezed) and by the red dots (the case it is finitely squeezed state with the same squeezing level as the input GKP state). The blue and green lines show the case where cubic phase gate is used as \cite{gottesman2001encoding, hastrup2021cubic} and the case where it is used with optimization, respectively. }
\label{figfide}
\end{figure}

In order to evaluate the lower limit of the performance of our proposed method, we consider an equally superposed state where $a = b = \frac{1}{\sqrt{2}}$ as the input state \cite{PhysRevA.101.032315}. On the basis of approximate GKP qubits, Eqs.~(\ref{approximate0L}) and (\ref{approximate1L}), the states of the operation target "in" and the non-Gaussian ancilla "A" can be written as:
\begin{align}
\ket{\psi_\Delta}_\mathrm{in}  &= \frac{1}{\sqrt{2}}  \left( \ket{0_\Delta}_\mathrm{in} +  \ket{1_\Delta}_\mathrm{in} \right)  \label{Eq:inDelta}\\
\ket{T_\Delta}_\mathrm{A} &= \frac{1}{\sqrt{2}} \left( \ket{0_\Delta}_\mathrm{A} +  \mathrm{e}^{i \frac{\pi}{4}}\ket{1_\Delta}_\mathrm{A} \right) . \label{Eq:TDelta}
\end{align}  
Note that we consider the same squeezing for both states. The Gaussian ancilla in mode "B" is a squeezed vacuum state given by
\begin{eqnarray}
\ket{\mathrm{Sq}_\sigma}_\mathrm{B} \propto  \int_{-\infty}^{\infty} dx \exp{ \left[ -\frac{x^2}{2 \left( 2\sigma^2\right)} \right] }    \ket{x}_\mathrm{B} . \label{Eq:sqsigma}
\end{eqnarray}
We consider two cases; an approximate case where $\sigma = \Delta$, which means the same squeezing level as GKP states,  and a case $\sigma \to 0$, which corresponds to the ideal $\ket{x=0}_\mathrm{B}$. By using modular subsystem decomposition \cite{pantaleoni2020modular}, we can obtain the logical density matrix $\hat{\rho}_{\Delta, \sigma}^\mathcal{L}$ from the output sate $\hat{\rho}_{\Delta, \sigma}$ of the circuit in Fig.~\ref{fig3}, where $\mathcal{L}$ indicates a logical subsystem. $\ket{T}_\mathcal{L} =\frac{1}{\sqrt{2}} \left( \ket{0}_\mathcal{L} + \mathrm{e}^{i\frac{\pi}{4}} \ket{1}_\mathcal{L} \right)$ as the target state,  we calculate the logical fidelities
\begin{equation}
F^\mathcal{L}_{\Delta, \sigma} = \subscripts{\mathcal{L}}{\Braket{T| \hat{\rho}_{\Delta, \sigma}^{\mathcal{L}} |T}} {\mathcal{L}} .
\label{Eq:LogicalFidelity}
\end{equation}
This logical fidelity can be used as a figure of merit to evaluate the performance of the T gate. 
Note that the CV fidelity does not always correctly represent the similarity of logical information. For example, even if the output and target GKP states have the same logical information, the CV fidelity between them becomes low if they have different squeezing levels. We can avoid this problem by using modular subsystem decomposition because we can retrieve logical information from any CV state, regardless of the properties such as squeezing level. Also, note that the relationship between the logical fidelity of the T gate and error correctability in concatenated code using GKP qubits is not well-studied and  expected to be revealed in future works. See Appendix A and B for detailed calculations. 

In Fig.~\ref{figfide}, the fidelity is plotted as a function of the squeezing level of the GKP states. The fidelities of cases $\sigma = \Delta$ and $\sigma \to 0$ are plotted as red dots and an orange line, respectively. 
The fidelity for $\sigma = \Delta$ is worse than that for the ideal case with $\sigma \to 0$, but the effect is dominated by the dependence on $\Delta$. When $\Delta$ is sufficiently small, the effect of imperfect Gaussian ancilla is minimal. For example, when the squeezing level $\gtrsim$ 10dB, both scenarios overcome the logical fidelity of 0.90. Such a squeezing level is achieved for squeezed vacuum in an experiment \cite{PhysRevLett.117.110801}. Moreover, there is a proposal to generate the GKP state with the squeezing level of 10 dB using existing techniques \cite{PhysRevA.101.032315}. A higher squeezing level is expected to be achieved with the advancement of technology. Our setup is experimentally feasible since it can work well as the T gate even when considering physically achievable states for all ancillary and input states. Furthermore, compared to the QND-based circuit in Fig. \ref{fig1}(a), our proposed circuit has better fidelity since it requires two fewer ancillary states. For comparison, we also calculate the fidelity of the T gate using a cubic phase gate together with Gaussian gates, 
\begin{eqnarray}
\hat{U}_\mathrm{CPG}\mathalpha{=}\mathrm{exp} \! \left\{ 
i \pi \left[ c_0 \left( \frac{\hat{x}}{\sqrt{\pi}}\right)^3\! \mathalpha{+} c_1 \left( \frac{\hat{x}}{\sqrt{\pi}}\right)^2 \! \mathalpha{+} c_2 \frac{\hat{x}}{\sqrt{\pi}}\right]\right\}
\label{Eq:CPGmaintext}
\end{eqnarray}
where $c_0, c_1$, and $c_2$ are coefficients to determine the gains of each gate. In the GKP's original paper, the case of $c_0 = \frac{1}{2}, c_1 = \frac{1}{4}, c_2 = -\frac{1}{2}$ was proposed \cite{gottesman2001encoding}. The fidelity of this case is plotted as a blue line and it saturates $\sim$0.78 \cite{hastrup2021cubic}. We find that the fidelity can be improved by optimizing the gains (details about the optimization are given in Appendix C), and when $c_0 = -\frac{1}{6}, c_1 = \frac{1}{4}, c_2 = \frac{1}{6}$, the fidelity goes up to $\sim$ 0.95 (plotted as a green line). This improvement may allow us to use the cubic phase gate to achieve universality. A noisy magic state prepared by $\hat{U}_\mathrm{CPG}$ with optimized gains on GKP state $\frac{1}{\sqrt{2}} \left( \ket{0_\Delta} + \ket{1_\Delta} \right)$ could achieve the threshold (fidelity $> 0.853$) for magic state distillation \cite{PhysRevA.71.022316, reichardt2005quantum, campbell2010bound}. We can expect to obtain a higher-quality magic state by using the higher-level encoding. This result is important because it can be applied not only for optical fields but also for any other bosonic fields employing GKP encodings, such as the ion-trapped system and the microwave system. Note that the fidelity threshold mentioned above is for qubits and can be directly applied to GKP qubits if the squeezing level is infinite. However, detailed analysis considering the imperfection of finite squeezing is expected in further studies. In future work, we will investigate the possibility of magic state distillation with our optimized cubic phase gate. Also, note that here we consider the ideal cubic phase gate represented as Eq.~(\ref{Eq:CPGmaintext}), in other words, we do not take into account the non-ideality of ancillary states like a cubic phase state. Therefore, our T gate setup based on the gate teleportation method is superior to the cubic phase gate approach. It is due to the high versatility and adjustability of the linear optical scheme with nonlinear feedforward at Fig.~\ref{fig3}.

\section{Discussion and conclusion}
Our proposal shows that nonlinear feedforward is versatile and can be used not only for the cubic phase gate but also for the T gate. We expect that other logical gates on GKP qubits, if properly decomposed, can be constructed similarly to our scheme by applying the gate teleportation method with appropriate ancillary GKP states and nonlinear feedforward operations \cite{PhysRevA.62.052316}. In addition to the logical gates on GKP qubits, nonlinear feedforward is an important component of other types of non-Gaussian operations \cite{PhysRevA.97.022329}. The current nonlinear feedforward is programmable because it is implemented using the digital FPGA. Therefore, it can be broadly applied to various kinds of non-Gaussian operations with light and will become an indispensable technology for quantum processing with light. Our method of  implementing the T gate is based on a beam splitter coupling, and it is simpler than the direct implementation of the originally proposed circuit \cite{gottesman2001encoding} because of fewer ancillary states. In addition, our scheme profits from the magic state injection method and is thus compatible with theories that utilize the magic state in the GKP qubit encoding \cite{PhysRevLett.123.200502,yamasaki2020cost}. Moreover, we have analyzed the performance of the T gate when the input and ancillary states are approximate states and found that our scheme surpasses the limit given by  the T gate implemented by the cubic phase gate even when the gains are optimized. By using our versatile method, the non-Clifford gates can be fully optimized and realized, thus a road towards the fault-tolerant universal optical quantum computer using GKP qubits is open. 

\section*{Acknowledgments}
This work was partly supported by JST [Moonshot R\&D][Grant Number JPMJMS2064], JSPS KAKENHI (Grant No. 18H05207, No. 18H01149, and No. 21J11615), UTokyo Foundation, and donations from Nichia Corporation. P. M. acknowledges grant GA18-21285S of the Czech Science Foundation. P.M. and R.F. were also supported by  support by national funding from MEYS and European Union’s Horizon 2020 (2014–2020) research and innovation framework programme under grant agreement No. 731473 (project 8C20002 ShoQC). Project ShoQC has received funding from the QuantERA ERA-NET Cofund in Quantum Technologies implemented within the European Unions Horizon 2020 Programme. R.F. acknowledges the project 21-13265X of the Czech Science Foundation. This project has received funding from the European Union's Horizon 2020 research and innovation programme (CSA Twinning) under grant agreement 951737 (NONGAUSS).
\appendix

\section{Modular Subsystem Decomposition}
In order to analyze the performance of the T gate when using approximate GKP sates, we employ the modular subsystem decomposition introduced by Pantaleoni {\it et al.} \cite{pantaleoni2020modular}. Here we write a short review of it. Any eigenstate $\ket{x} (x \in \mathbb{R})$ of the quadrature operator $\hat{x}$ in CV Hilbert space $\mathcal{H}_\mathrm{CV}$ can be divided into the integer part $m$ and the fractional part $u$ modulo $\sqrt{\pi}$ as 
\begin{align}
\ket{x} = \ket{\sqrt{\pi} m + u }  =: \ket{m, u}
\end{align}
where 
\begin{gather}
m = \left \lfloor \frac{x}{\sqrt{\pi}} + \frac{1}{2} \right \rfloor \in \mathbb{Z} \\
u = x - \sqrt{\pi} m \in \left[ -\frac{\sqrt{\pi}}{2}, \frac{\sqrt{\pi}}{2} \right)
\end{gather}
where $\left \lfloor \cdot \right \rfloor $ is the floor function. We can decompose $\mathcal{H}_\mathrm{CV}$ into two subsystems as $\mathcal{L} \otimes \mathcal{G}$ by using
\begin{align}
\ket{x} = \ket{m, u} =: \ket{l}_\mathcal{L} \otimes \ket{ \tilde{m}, \tilde{u}}_\mathcal{G}
\end{align}
where 
\begin{gather}
l \equiv m~(\mathrm{mod~2}) \in  \{0, 1\} \\
\tilde{m} = \frac{1}{2} (m - l) \in \mathbb{Z} \label{suppleEq:gaugeorigin} \\
\tilde{u} = u \in \left[ -\frac{\sqrt{\pi}}{2}, \frac{\sqrt{\pi}}{2} \right). 
\end{gather}
$\mathcal{L} \cong \mathbb{C}^2$ represents the logical qubit space and $\mathcal{G}$ is the
remaining CV gauge mode. We can obtain the logical density matrix $\hat{\rho}^\mathcal{L}$ from any CV state $\hat{\rho}$ including a mixed state by tracing over the gauge mode as
\begin{align}
\hat{\rho}^\mathcal{L} = \mathrm{Tr}_\mathcal{G}\left[ \hat{\rho}\right] .
\label{Eq:modular}
\end{align}

\begin{widetext}
\section{details of fidelity calculation}
Here we describe how we calculated the logical fidelity of our T gate in Fig.~\ref{figfide} in the main text. We first calculate the output state of  our T gate shown in Fig.~\ref{fig3}. In the case of finite squeezing levels, the ancillary T state and the input state are 
\begin{align}
\ket{T_\Delta}_\mathrm{A} &=  \frac{1}{\sqrt{2}}  \left( \ket{0_\Delta}_\mathrm{A} + \mathrm{e}^{i \frac{\pi}{4}} \ket{1_\Delta}_\mathrm{A} \right) \\
\ket{\psi_\Delta}_\mathrm{in} &=  a \ket{0_\Delta}_\mathrm{in} + b \ket{1_\Delta}_\mathrm{in} .
\end{align}
$\ket{0_\Delta}, \ket{1_\Delta}$ are approximate GKP states defined as Eqs.~(\ref{approximate0L}) and (\ref{approximate1L}). The wave functions of these states in position basis are represented as 
\begin{alignat}{2}
\ket{T_\Delta}_\mathrm{A} &=&  \int_{-\infty}^{\infty} dx_1 \subscripts{\mathrm{A}}{\braket{x_1 |T_\Delta }}{\mathrm{A}} \ket{x_1}_\mathrm{A} &= N_{T, \Delta}  \int_{-\infty}^{\infty} dx_1 T_\Delta(x_1) \ket{x_1}_\mathrm{A} \\
\ket{\psi_\Delta}_\mathrm{in} &=& \int_{-\infty}^{\infty} dx_2 \subscripts{\mathrm{in}}{\braket{x_2 |\psi_\Delta }}{\mathrm{in}} \ket{x_2}_\mathrm{in} &= N_{\psi, \Delta}  \int_{-\infty}^{\infty} dx_2 \psi_\Delta(x_2) \ket{x_2}_\mathrm{in} 
\end{alignat}
where $N_{T, \Delta}, N_{\psi, \Delta}$ are normalization constants. We also consider a finitely squeezed vacuum for the ancilla in the mode ``B'', 
\begin{eqnarray}
\ket{\mathrm{Sq}_\sigma}_\mathrm{B} = N_{Sq, \sigma}  \int_{-\infty}^{\infty} dx_3 Sq_\sigma(x_3) \ket{x_3}_\mathrm{B} = N_{Sq, \sigma}  \int_{-\infty}^{\infty} dx_3 \exp{ \left[ -\frac{x_3^2}{2 \left( 2\sigma^2\right)} \right] }    \ket{x_3}_\mathrm{B}
\end{eqnarray}
where $\sigma^2$ is the variance of quadrature $x$ and $N_{Sq, \sigma}$ is a normalization constant. The squeezing level for this mode is $-10\log_{10} 2\sigma^2$.

The transformation by the first beam splitter is as follows:
\begin{align}
\ket{T_\Delta}_\mathrm{A} \ket{\psi_\Delta}_\mathrm{in} \ket{\mathrm{Sq}_\sigma}_\mathrm{B} = &N_{T, \Delta}  N_{\psi, \Delta} N_{Sq, \sigma}  \int_{-\infty}^{\infty} dx_1 \int_{-\infty}^{\infty} dx_2 \int_{-\infty}^{\infty} dx_3  T_\Delta(x_1) \psi_\Delta(x_2) Sq_\sigma(x_3) \ket{x_1}_\mathrm{A} \ket{x_2}_\mathrm{in} \ket{x_3}_\mathrm{B} \nonumber \\
\xrightarrow{\mathrm{BS1}} &N_{T, \Delta}  N_{\psi, \Delta} N_{Sq, \sigma}  \int_{-\infty}^{\infty} dx_1 \int_{-\infty}^{\infty} dx_2 \int_{-\infty}^{\infty} dx_3  T_\Delta(x_1) \psi_\Delta(x_2) Sq_\sigma(x_3) \Ket{\frac{- x_1 + x_2}{\sqrt{2}} }_\mathrm{A} \Ket{\frac{x_1 + x_2}{\sqrt{2}}}_\mathrm{in} \ket{x_3}_\mathrm{B}. 
\end{align}
In the first homodyne detection, we measure the $x$ quadrature of the mode ``A'' and obtain value $q_1$ with probability density $P_1(q_1)$.  By performing integration over $x_1$ using the following formula, 
\begin{equation}
\subscripts{\mathrm{A}}{\Braket{q_1| \frac{- x_1 + x_2}{\sqrt{2} } }}{ \mathrm{A}}  = \sqrt{2} \delta(\sqrt{2} q_1 + x_1 - x_2 ) 
\end{equation}
the normalized state after the first homodyne detection becomes 
\begin{equation}
\frac{\sqrt{2} N_{T, \Delta}  N_{\psi, \Delta} N_{Sq, \sigma}}{\sqrt{P_1(q_1)}} \int_{-\infty}^{\infty} dx_2 \int_{-\infty}^{\infty} dx_3  T_\Delta(x_2 - \sqrt{2} q_1) \psi_\Delta(x_2) Sq_\sigma(x_3) \Ket{\sqrt{2} x_2 - q_1}_\mathrm{in} \ket{x_3}_\mathrm{B}  .
\end{equation}
After the second beam splitter, this state becomes
\begin{equation}
\xrightarrow{\mathrm{BS2}} \frac{\sqrt{2} N_{T, \Delta}  N_{\psi, \Delta} N_{Sq, \sigma}}{\sqrt{P_1(q_1)}}  \int_{-\infty}^{\infty} dx_2 \int_{-\infty}^{\infty} dx_3  T_\Delta(x_2 - \sqrt{2} q_1) \psi_\Delta(x_2) Sq_\sigma(x_3) \Ket{x_2 - \frac{q_1 - x_3}{\sqrt{2}} }_\mathrm{in} \Ket{x_2 - \frac{q_1 + x_3}{\sqrt{2}}}_\mathrm{B}.
\end{equation}
Then we apply phase rotation according to $q_1$ on the mode ``in''. When using ideal GKP states, $\sqrt{\frac{2}{\pi}} q_1$ takes an only integer value. However, we consider approximate GKP states in which each comb has a finite spread, thus $\sqrt{\frac{2}{\pi}} q_1$ does not always take an integer value. Here we consider the nearest inter value of $\sqrt{\frac{2}{\pi}} q_1$ and according to its evenness, the $\kappa$ is defined as    
\begin{eqnarray}
\kappa (q_1)  \equiv \left \lfloor \sqrt{\frac{2}{\pi} } q_1 + \frac{1}{2}  \right \rfloor (\mathrm{mod~2}), 
\label{Eq:kappa}
\end{eqnarray}
which takes a discrete value 0 or 1. Using this $\kappa(q_1)$, the rotation angle is defined as in the main text, 
\begin{equation}
\theta\bigl(\kappa\left(q_1\right)\bigr) = \arctan\left[\kappa\left(q_1\right)\right].
\end{equation}
After applying $\hat{R}_\mathrm{in} \Bigl(-\theta\bigl(\kappa(q_1) \bigr)\Bigr)$, we measure the $p$ quadrature value of the mode ``in'' at the second homodyne detection and obtain value $q_2$. We define conditional probability density $P_2(q_2|q_1)$ of taking $q_2$ in the second homodyne detection when the measurement value of the first homodyne detection is $q_1$. Note that the joint  probability density $P(q_1, q_2)$ is written as 
\begin{equation}
P(q_1, q_2) = P_1(q_1) P_2(q_2|q_1) .
\end{equation}
Using $\subscripts{\mathrm{in} }{\bra{p = q_2}}{ } = \subscripts{\mathrm{in} } {\bra{q_2}} { } \hat{R}_\mathrm{in} \left(\frac{\pi}{2}\right)$, the normalized state after the second homodyne detection becomes
\begin{align}
&\quad \frac{\sqrt{2} N_{T, \Delta}  N_{\psi, \Delta} N_{Sq, \sigma} }{\sqrt{P_1(q_1) P_2(q_2|q_1)} } \int_{-\infty}^{\infty} \! dx_2 \! \int_{-\infty}^{\infty} \! dx_3  T_\Delta(x_2 \mathalpha{-} \sqrt{2} q_1) \psi_\Delta(x_2) Sq_\sigma(x_3)  \subscripts{\mathrm{in}} {\Braket{q_2 | \hat{R}_\mathrm{in}\left( \frac{\pi}{2} \mathalpha{-} \theta\bigl(\kappa\left(q_1\right)\bigr) \right) |x_2 \mathalpha{-} \frac{q_1 \mathalpha{-} x_3}{\sqrt{2}} }}{\mathrm{in}} \Ket{x_2 \mathalpha{-} \frac{q_1 \mathalpha{+} x_3}{\sqrt{2}}}_\mathrm{B} \nonumber \\ 
&= \frac{N_{T, \Delta}  N_{\psi, \Delta} N_{Sq, \sigma}  \left(1+ \kappa\left(q_1\right)^2\right)^{\frac{1}{4} }} { \sqrt{\pi P(q_1, q_2) }} \int_{-\infty}^{\infty} dx_2 \int_{-\infty}^{\infty} dx_3 T_\Delta(x_2 - \sqrt{2} q_1) \psi_\Delta(x_2) Sq_\sigma(x_3)  \nonumber \\ 
&\hspace{25mm}  \times \exp{ \left\{ i         \left [   \left( q_2^2 + \left( x_2 - \frac{q_1-x_3 }{\sqrt{2} } \right)^2   \right) \frac{\kappa(q_1) }{2}  -q_2 \left( x_2- \frac{q_1-x_3 }{\sqrt{2}}\right)\sqrt{ 1+\kappa(q_1)^2}     \right ] \right\} }   \Ket{x_2 - \frac{q_1 + x_3}{\sqrt{2}}}_\mathrm{B} .
\end{align}
Here we used  
\begin{equation}
\braket{x | \hat{R}(\Theta) | x^{\prime}} = \frac{1}{\sqrt{2\pi |\sin \Theta | } } \exp \left\{ \frac{i \left[ (x^2 + { x^{\prime}}^2) \cos \Theta - 2xx^{\prime} \right] }{2 \sin \Theta} \right\} , 
\end{equation}
the definition of $\kappa(q_1)$ in Eq.~(\ref{Eq:kappa}), and the formula derived from it:
\begin{equation}
\frac{1}{\cos \theta (\kappa(q_1))} = \sqrt{1+\kappa(q_1)^2}. 
\end{equation}
Finally by performing the displacement operation on the mode ``B'' 
\begin{equation}
\hat{D}_\mathrm{B} \left( \frac{q_1}{\sqrt{2}},  \kappa(q_1) \frac{q_1}{\sqrt{2}} + \sqrt{1 + \kappa(q_1)^2} q_2 \right) ,
\end{equation}
we obtain the output of our T gate
\begin{align}
&\ket{\psi_{\Delta, \sigma}^{\prime} (q_1, q_2) }_\mathrm{out}
\mathalpha{=}  \exp \left[ \frac{i}{2} \left( \sqrt{\frac{ 1\mathalpha{+}\kappa(q_1)^2}{2} }q_1 q_2 \mathalpha{+} \kappa(q_1) q_2^2 \right) \right] \frac{N_{T, \Delta}  N_{\psi, \Delta} N_{Sq, \sigma}  \Bigl(1\mathalpha{+} \kappa(q_1)^2\Bigr)^{\frac{1}{4} }} { \sqrt{\pi P(q_1, q_2)} } \nonumber \\
&\mathalpha{\times} \! \int_{-\infty}^{\infty} \! dx_2 \! \int_{-\infty}^{\infty} \! dx_3 
 T_\Delta(x_2 \mathalpha{-} \sqrt{2} q_1) \psi_\Delta(x_2) Sq_\sigma(x_3) 
\exp{ \left\{ i \left[  \kappa(q_1) \left( 
\frac{1}{2} \left( x_2 \mathalpha{+} \frac{x_3}{\sqrt{2} } \right)^2
 \mathalpha{-}q_1x_3 \right) \mathalpha{-} \sqrt{2\Bigl(1 \mathalpha{+}\kappa(q_1)^2\Bigr)} q_2 x_3   \right]  \right\} }   \Ket{x_2 \mathalpha{-} \frac{x_3}{\sqrt{2}}}_\mathrm{out}.
\end{align}
The output state is a mixed state conditioned by the measurement values $q_1$ and $q_2$, therefore the density matrix can be written as  
\begin{equation}
\hat{\rho}_{\Delta, \sigma} = \int_{-\infty}^\infty dq_1 \int_{-\infty}^\infty dq_2 P(q_1, q_2) \ket{\psi_{\Delta, \sigma}^\prime (q_1, q_2)  }_\mathrm{out}  \subscripts{\mathrm{out}}{\bra{\psi_{\Delta, \sigma}^\prime (q_1, q_2)  }}{}
\label{Eq:outdensity}
\end{equation}

Next, we calculate the logical density matrix. By inserting Eq.~(\ref{Eq:outdensity}) into Eq.~(\ref{Eq:modular}), the $(\zeta, \eta )$ component of the density matrix can be calculated as ($\zeta, \eta = 0, 1$)
\begin{align}
&\quad \subscripts{\mathcal{L} }{ \Braket{\zeta | \hat{\rho}_{\Delta, \sigma}^{\mathcal{L}} | \eta}}{ \mathcal{L}} \nonumber \\
&= \sum_{\tilde{m} \in \mathbb{Z}} \int_{-\frac{\sqrt{\pi}}{2} }^{\frac{\sqrt{\pi}}{2}} d\tilde{u}~\biggl( \subscripts{\mathcal{L} }{\Bra{\zeta} }{} \otimes \subscripts{\mathcal{G} }{\Bra{\tilde{m}, \tilde{u}} }{}  \biggr) \hat{\rho}_{\Delta, \sigma} \biggl( \Ket{\eta}_{\mathcal{L}} \otimes \Ket{\tilde{m}, \tilde{u}}_{\mathcal{G}} \biggr) \nonumber \\
&= \sum_{\tilde{m} \in \mathbb{Z}} \int_{-\frac{\sqrt{\pi}}{2} }^{\frac{\sqrt{\pi}}{2}} d\tilde{u} \int_{-\infty}^{\infty}dq_1 \int_{-\infty}^{\infty}dq_2~ P(q_1, q_2)   \subscripts{\mathrm{out}}{ \Braket{ \left( 2\tilde{m} + \zeta \right) \sqrt{\pi} + \tilde{u}  |\psi_{\Delta, \sigma}^{\prime} (q_1, q_2) }}{\mathrm{out}} \subscripts{\mathrm{out}}{ \Braket{  \psi_{\Delta, \sigma}^{\prime} (q_1, q_2)  | \left( 2\tilde{m} + \eta \right) \sqrt{\pi} + \tilde{u} }}{\mathrm{out}} \nonumber \\
&= \frac{\left| N_{T, \Delta}  N_{\psi, \Delta} N_{Sq, \sigma} \right|^2  }{\pi} \sum_{\tilde{m} \in \mathbb{Z}} \int_{-\frac{\sqrt{\pi}}{2} }^{\frac{\sqrt{\pi}}{2}} d\tilde{u} \int_{-\infty}^{\infty}dq_1 \int_{-\infty}^{\infty}dq_2
\int_{-\infty}^{\infty} dx_2 \int_{-\infty}^{\infty} dx_3 \int_{-\infty}^{\infty} dx_2^\prime \int_{-\infty}^{\infty}  dx_3^\prime~\sqrt{1 + \kappa \left( q_1 \right)^2} \nonumber \\
&\hspace{10mm} \times T_\Delta(x_2 - \sqrt{2} q_1) \psi_\Delta(x_2) Sq_\sigma(x_3) 
\exp{ \left\{ i \left[  \kappa(q_1) \left( \frac{1}{2} \left( x_2 \mathalpha{+} \frac{x_3}{\sqrt{2} } \right)^2 -q_1x_3 \right) - \sqrt{2\Bigl(1 + \kappa(q_1)^2\Bigr)} q_2 x_3   \right]  \right\} } \nonumber \\
&\hspace{10mm}\times T_\Delta^*(x_2^\prime - \sqrt{2} q_1) \psi_\Delta^*(x_2^\prime) Sq_\sigma^*(x_3^\prime) 
\exp{ \left\{ -i \left[  \kappa(q_1) \left(  \frac{1}{2} \left( x_2^{\prime} \mathalpha{+} \frac{x_3^{\prime}}{\sqrt{2} } \right)^2 - q_1x_3^\prime \right) - \sqrt{2\Bigl(1 +\kappa(q_1)^2\Bigr)} q_2 x_3^\prime   \right]  \right\} } \nonumber \\
&\hspace{60mm}\times \delta\left( x_2 - \frac{x_3}{\sqrt{2}} - \left( 2 \tilde{m} + \zeta \right) \sqrt{\pi} - \tilde{u} \right) 
\delta\left( x_2^\prime - \frac{x_3^\prime}{\sqrt{2}} - \left( 2 \tilde{m} + \eta \right) \sqrt{\pi} - \tilde{u} \right).
\label{Eq:modular2}
\end{align}
The integral for $q_2$ can be performed as
\begin{align}
\int_{-\infty}^\infty dq_2 \exp \left[ -i \sqrt{2\left( 1 + \kappa \left(  q_1\right)^2\right) } q_2 (x_3 - x_3^\prime) \right] = \pi \sqrt{ \frac{2}{1 + \kappa \left(  q_1\right)^2 } } \delta(x_3 - x_3^\prime), 
\end{align}
and performing the integral for $x_2, x_2^\prime, x_3^\prime$, Eq.~(\ref{Eq:modular2}) can be calculated as   
\begin{align}
&\quad \subscripts{\mathcal{L} }{ \Braket{\zeta | \hat{\rho}_{\Delta, \sigma}^{\mathcal{L}} | \eta}}{ \mathcal{L}} 
= \sqrt{2}  \left| N_{T, \Delta}  N_{\psi, \Delta} N_{Sq, \sigma} \right|^2  \sum_{\tilde{m} \in \mathbb{Z}} \int_{-\frac{\sqrt{\pi}}{2} }^{\frac{\sqrt{\pi}}{2}} d\tilde{u} \int_{-\infty}^{\infty}dq_1\int_{-\infty}^{\infty} dx_3 \left|Sq_\sigma(x_3)\right|^2 \nonumber \\  
&\times T_\Delta \left(  \frac{x_3}{\sqrt{2}} + \left( 2 \tilde{m} + \zeta \right) \sqrt{\pi} + \tilde{u}  - \sqrt{2} q_1 \right)  \psi_\Delta \left( \frac{x_3}{\sqrt{2}} + \left( 2 \tilde{m} + \zeta \right) \sqrt{\pi} + \tilde{u} \right)  \exp \left\{ \frac{i}{2} \kappa\left( q_1\right)  \left[\sqrt{2} x_3 + \left( 2 \tilde{m} + \zeta \right) \sqrt{\pi} + \tilde{u} \right]^2 \right\} \nonumber \\
&\times T_\Delta^* \left(  \frac{x_3}{\sqrt{2}} + \left( 2 \tilde{m} + \eta \right) \sqrt{\pi} + \tilde{u}  - \sqrt{2} q_1 \right)  \psi_\Delta^* \left( \frac{x_3}{\sqrt{2}} + \left( 2 \tilde{m} + \eta \right) \sqrt{\pi} + \tilde{u} \right) \exp \left\{ -\frac{i}{2} \kappa\left( q_1\right)  \left[\sqrt{2} x_3 + \left( 2 \tilde{m} + \eta \right) \sqrt{\pi} + \tilde{u} \right]^2  \right\} .
\label{Eq:result_finite}
\end{align}
Here we consider the case where $a = b = 1$ as an input state. The target state is $\ket{T}_\mathcal{L} = \ket{0}_\mathcal{L} +  \mathrm{e}^{i \frac{\pi}{4}} \ket{1}_\mathcal{L}$, therefore the logical fidelity is
\begin{equation}
F^\mathcal{L}_{\Delta, \sigma} = \subscripts{\mathcal{L}}{\Braket{T| \hat{\rho}_{\Delta, \sigma}^{\mathcal{L}} |T}} {\mathcal{L}} .
\label{Eq:fide_finite}
\end{equation}
For various squeezing levels, we calculate numerically Eq.~(\ref{Eq:fide_finite}) when  $\sigma = \Delta$. The result is plotted as the red dots in Fig.~\ref{figfide} in the main text. For programming, we reference \cite{PhysRevA.101.032315}.

When the ancillary squeezed vacuum is ideal i.e. $\sigma \to 0$,  we can simplify Eq.~(\ref{Eq:result_finite}). By using the following relation
\begin{align}
\lim_{\sigma \to 0}  \left| N_{Sq, \sigma}  Sq_\sigma(x_3) \right|^2 = \delta(x_3), 
\end{align}
Eq.~(\ref{Eq:result_finite}) can be calculated as
\begin{align}
&\quad \subscripts{\mathcal{L} }{ \Braket{\zeta | \hat{\rho}_{\Delta, \sigma \to 0}^{\mathcal{L}} | \eta}}{ \mathcal{L}} \nonumber \\ 
&\mathalpha{=} \sqrt{2}  \left| N_{T, \Delta}  N_{\psi, \Delta} \right|^2  \! \sum_{\tilde{m} \in \mathbb{Z}} \! \int_{-\frac{\sqrt{\pi}}{2} }^{\frac{\sqrt{\pi}}{2}} \! d\tilde{u} \! \int_{-\infty}^{\infty} \! dq_1  
 T_\Delta \left( \left( 2 \tilde{m} \mathalpha{+} \zeta \right) \sqrt{\pi} \mathalpha{+} \tilde{u}  \mathalpha{-} \sqrt{2} q_1 \right)  \psi_\Delta \Bigl( \left( 2 \tilde{m} \mathalpha{+} \zeta \right) \sqrt{\pi} \mathalpha{+} \tilde{u} \Bigr) \exp \left\{ \frac{i}{2} \kappa\left( q_1\right)  \left[ \left( 2 \tilde{m} \mathalpha{+} \zeta \right) \sqrt{\pi} \mathalpha{+} \tilde{u} \right]^2  \right\} \nonumber \\
&\hspace{35mm} \times T_\Delta^* \left(  \left( 2 \tilde{m} \mathalpha{+} \eta \right) \sqrt{\pi} \mathalpha{+} \tilde{u}  \mathalpha{-} \sqrt{2} q_1 \right)  \psi_\Delta^* \Bigl( \left( 2 \tilde{m} \mathalpha{+} \eta \right) \sqrt{\pi} \mathalpha{+} \tilde{u} \Bigr)  \exp \left\{ \mathalpha{-}\frac{i}{2} \kappa\left( q_1\right)  \left[ \left( 2 \tilde{m} \mathalpha{+} \eta \right) \sqrt{\pi} \mathalpha{+} \tilde{u} \right]^2 \right\}. 
\label{Eq:result_infinite}
\end{align}
Logical fidelity $F^\mathcal{L}_{\Delta, \sigma \to 0}$ can be obtained by using Eq.~(\ref{Eq:fide_finite}) as well. The result is plotted as the orange line in Fig.~\ref{figfide} in the main text. 

\end{widetext}
\section{optimization of the cubic phase gate as the T gate}
\begin{figure}
\centering \includegraphics[angle=0, scale=0.53]{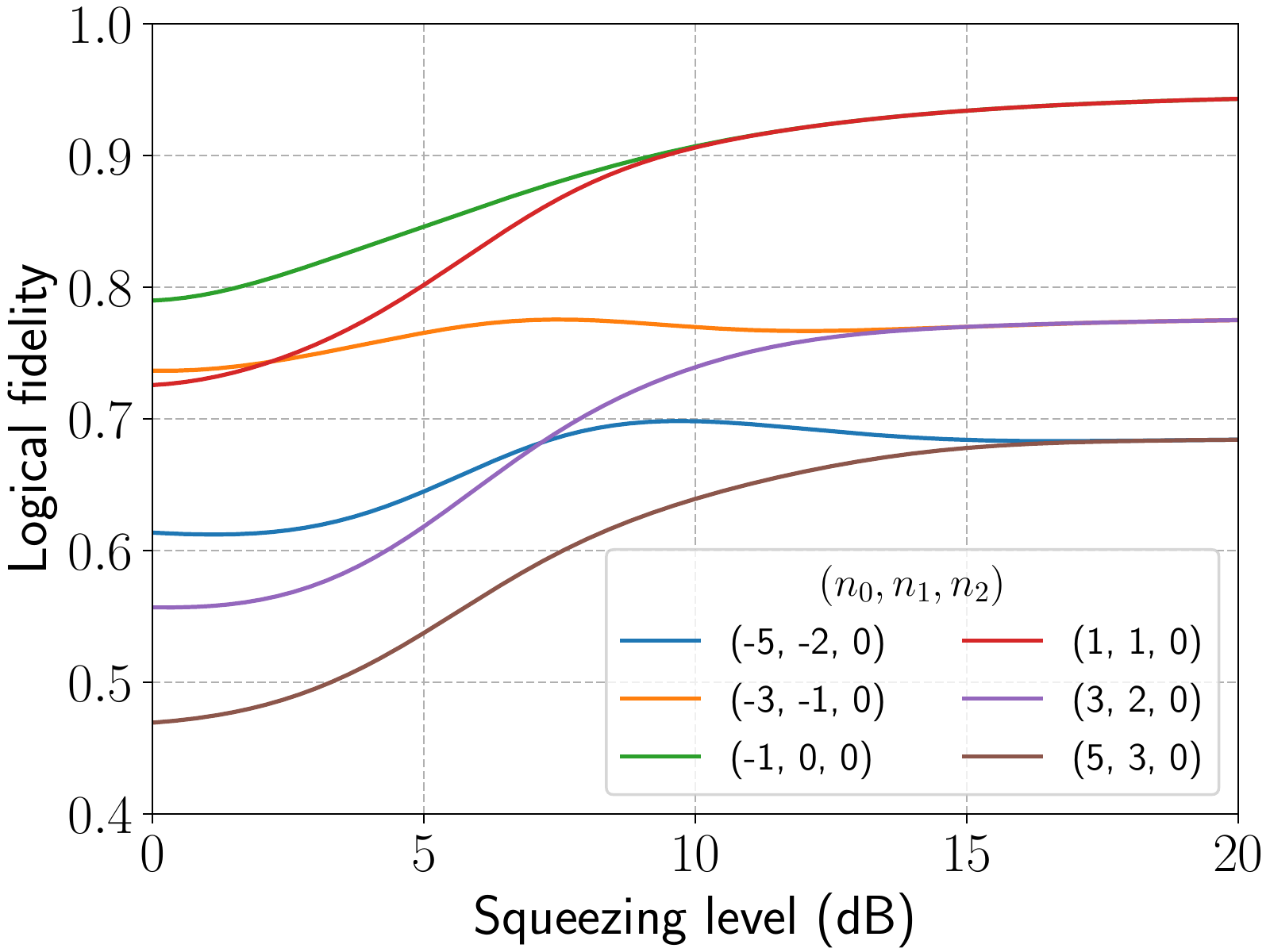} 
\caption{The logical fidelity as the T gate when the gains of cubic phase gate and other Gaussian gates are ajusted.
}
\label{Fig:CPG_supple}
\end{figure}

In general, there are many physical gates corresponding to a certain logical gate on GKP qubits. A simple example is the logical X gate, which can be realized by $\hat{D}\left( (2n+1)\sqrt{\pi}, 0 \right)$ for any $n \in \mathbb{Z}$. The same is true for the logical T gate. In the original paper \cite{gottesman2001encoding}, GKP proposed the implementation of the T gate by combining a single cubic phase gate with Gaussian operations:
\begin{align}
\hat{U} \mathalpha{=} \mathrm{exp}  \left\{ 
i \pi \left[ \frac{1}{2} \left( \frac{\hat{x}}{\sqrt{\pi}}\right)^3 \mathalpha{+} \frac{1}{4} \left( \frac{\hat{x}}{\sqrt{\pi}}\right)^2  \mathalpha{-} \frac{1}{2} \frac{\hat{x}}{\sqrt{\pi}}\right]\right\},
\label{Eq:CPGGKP}
\end{align}
which was pointed out not to be suitable for the T gate \cite{hastrup2021cubic}. However, we can think of other physical implementations of the T gate which has better performance even if only one cubic phase gate is used. Here we consider 
\begin{align}
\hat{U}_\mathrm{CPG} \mathalpha{=} \mathrm{exp}  \left\{ 
i \pi \left[ c_0 \left( \frac{\hat{x}}{\sqrt{\pi}}\right)^3 \mathalpha{+} c_1 \left( \frac{\hat{x}}{\sqrt{\pi}}\right)^2 \mathalpha{+} c_2 \frac{\hat{x}}{\sqrt{\pi}}\right]\right\}, 
\label{Eq:CGPGeneral}
\end{align}
which is a generalized form of Eq.~(\ref{Eq:CPGGKP}). $(c_0, c_1,c_2)$ are gains of each physical gate.  We first consider the conditions under which Eq.~(\ref{Eq:CGPGeneral}) works as the T gate on ideal GKP qubits. For any input state $a \ket{0_\mathrm{L}} + b \ket{1_\mathrm{L}}$, the result of acting $\hat{U}_\mathrm{CPG}$ is 
\begin{align}
 &\hat{U}_\mathrm{CPG} \left( a \ket{0_\mathrm{L}} + b \ket{1_\mathrm{L}} \right) \nonumber \\
=& a \sum_{s \in \mathbb{Z}} \mathrm{e}^{ 2i\pi f(s) } \ket{2s\sqrt{\pi}} + b \sum_{s \in \mathbb{Z}} \mathrm{e}^{ 2i\pi f(s)}  \mathrm{e}^{2i\pi g(s)} \mathrm{e}^{ i\pi h} \ket{(2s+1) \sqrt{\pi}}
\label{Eq:CPGout}
\end{align}
where
\begin{align}
f(s) &= 4 c_0 s^3 + 2c_1 s^2 + c_2 s  \\
g(s)&= 6 c_0 s^2 + ( 3 c_0 + 2 c_1) s  \\
h &= c_0 + c_1 + c_2 
\end{align}
The necessary and sufficient conditions for Eq.~(\ref{Eq:CPGout}) to match the output state of the T gate $a \sum_{s \in \mathbb{Z}} \ket{2s\sqrt{\pi}} + \mathrm{e}^{i\frac{\pi}{4}} b \sum_{s \in \mathbb{Z}}\ket{(2s+1) \sqrt{\pi}}$ is 
\begin{align}
\mathrm{e}^{2i\pi f(s)} &= A  ~ (\forall s \in \mathbb{Z}) \\
\mathrm{e}^{2i\pi g(s)} e^{i\pi h} &= \mathrm{e}^{i\frac{\pi}{4}} ~  (\forall s \in \mathbb{Z}),
\end{align}
where $A$ is a constant. Since $f(0) = g(0) = 0$, these conditions can be rewritten as 
\begin{align}
\mathrm{e}^{2i\pi f(s)} &= 1  ~ (\forall s \in \mathbb{Z}) \\
\mathrm{e}^{2i\pi g(s)}&= 1  ~  (\forall s \in \mathbb{Z}) \\
e^{i\pi h} &= \mathrm{e}^{i\frac{\pi}{4}} .
\end{align}
Each condition can be transformed equivalently as follows:
\begin{align}
\mathrm{e}^{2i\pi f(s)} = 1~ (\forall s \in \mathbb{Z}) & \Leftrightarrow f(s) \in \mathbb{Z}~ (\forall s \in \mathbb{Z}) \nonumber \\
&\Leftrightarrow
\begin{cases}
f(-1)= -4c_0 + 2c_1 -c_2 \in \mathbb{Z} \\  
f(1)= 4c_0 + 2c_1 +c_2 \in \mathbb{Z} \\
f(2)= 32c_0 + 8c_1 + 2c_2 \in \mathbb{Z}
\end{cases} \nonumber \\
&\Leftrightarrow
\begin{cases}
4c_1 \in \mathbb{Z} \\  
4c_0 + 2c_1 +c_2 \in \mathbb{Z} \\
24c_0 \in \mathbb{Z} .
\end{cases}
\label{Eq:condition1}\\
\mathrm{e}^{2i\pi g(s)}= 1 ~  (\forall s \in \mathbb{Z}) &\Leftrightarrow g(s) \in \mathbb{Z} ~ (\forall s \in \mathbb{Z}) \nonumber \\
&\Leftrightarrow
\begin{cases}
g(1) = 9c_0 + 2c_1 \in \mathbb{Z}\\
g(-1) = 3c_0 - 2c_1 \in \mathbb{Z}
\end{cases} \nonumber \\
&\Leftrightarrow
\begin{cases}
9c_0 + 2c_1\in \mathbb{Z}\\
12c_0 \in \mathbb{Z}
\end{cases}
\label{Eq:condition2}\\
e^{i\pi h} = \mathrm{e}^{i\frac{\pi}{4}} &\Leftrightarrow 4h \equiv 1~\mathrm{(mod~8)} \nonumber \\
&\Leftrightarrow 4 (c_0 + c_1 + c_2) \equiv 1 ~\mathrm{(mod~8)}. \label{Eq:condition3}
\end{align}
By combining Eqs.~(\ref{Eq:condition1}), (\ref{Eq:condition2}), and (\ref{Eq:condition3}), we obtain 
\begin{align}
\begin{cases}
6c_0 \in \mathbb{Z} \\
3c_0 + 2c_1 \in \mathbb{Z} \\
c_0 + c_2 \in \mathbb{Z} \\
4 (c_0 + c_1 + c_2) \equiv 1 ~ \mathrm{(mod~8)}. \label{Eq:condition4}
\end{cases}
\end{align}
We define $6c_0 = n_0, ~3 c_0 + 2c_1 = n_1$, and $c_0+c_2 = n_2 $ where $(n_0, n_1,n_2)$ are integers. The fourth condition of Eq.~(\ref{Eq:condition4}) becomes 
\begin{align}
4 (c_0 + c_1 + c_2)  = -n_0 + 2 n_1 + 4n_2 \equiv 1~ \mathrm{(mod~8)}.
\end{align}
In summary, the conditions imposed on $(c_0, c_1, c_2)$ are
\begin{align}
\begin{cases}
c_0 = \frac{1}{6}n_0 \\
c_1 = -\frac{1}{4}n_0 + \frac{1}{2}n_1\\
c_2 = n_2 - \frac{1}{6}n_0 \\
-n_0 + 2 n_1 + 4n_2 \equiv 1~\mathrm{(mod~8)} .
\end{cases} \label{Eq:conditionLast}
\end{align}
GKP's original suggestion (Eq.~(\ref{Eq:CPGGKP})) corresponds to the case where $(n_0, n_1, n_2) = (3, 2,0)$. For various $(n_0, n_1, n_2)$ which satisfy Eq.~(\ref{Eq:conditionLast}), we calculate the logical fidelity as the T gate when using the approximate GKP state as the input in the same way as described before. The result is shown in Fig.~\ref{Fig:CPG_supple}. For each $n_0$, we choose and plot the case of $n_1$ where the logical fidelity is maximized (actually, the value of $n_2$ did not significantly affect the logical fidelity). In the high squeezing region, the smaller the absolute value of $n_0$, the higher the logical fidelity. This trend can be explained in the same way as discussed in \cite{hastrup2021cubic}. The approximate GKP states have a finite width at each peak, so the phase fluctuations caused by the cubic phase gate within that width make the T gate perform poorly. The smaller the absolute value of the gain of the cubic phase gate, the smaller the phase fluctuations  and thus the higher the logical fidelity. $n_0 = 0$ can not satisfy the conditions of Eq.~(\ref{Eq:conditionLast}), so $n_0 = 1$ and $-1$ become the optimal cases for the T gate. The optimum case $(n_0, n_1, n_2) = (-1, 0, 0)$, i.e. $(c_0, c_1, c_2) = (-\frac{1}{6},\frac{1}{4}, \frac{1}{6})$ is also plotted in Fig.~\ref{figfide} in the main text as the green line. Where the squeezing level is small, the behavior is different for positive and negative $n_0$. This is due to the asymmetry of the gauge mode corresponding to the logical subsystem $\ket{1}_\mathcal{L}$ in the modular subsystem decomposition. For example, the origin of that gauge mode $\ket{1}_\mathcal{L} \otimes \ket{0,0}_\mathcal{G}$ corresponds to $\ket{\sqrt{\pi}}$ in the original Hilbert space $\mathcal{H}_\mathrm{CV}$. If we take this origin to $\ket{-\sqrt{\pi}}$, which means to make the sign of $l$ positive in Eq.~(\ref{suppleEq:gaugeorigin}), the behavior of fidelity in Fig.~(\ref{Fig:CPG_supple}) is reversed for positive and negative $n_0$.

\bibliography{ref.bib}

\end{document}